# Investigation of Partition Cells as a Structural Basis Suitable for Assessments of Individual Scientists[1]

Nadine Rons[*]

[*]*Nadine.Rons@vub.ac.be*
Research Coordination Unit, Vrije Universiteit Brussel (VUB), Pleinlaan 2, B-1050 Brussels (Belgium)

**Abstract**
Individual, excellent scientists have become increasingly important in the research funding landscape. Accurate bibliometric measures of an individual's performance could help identify excellent scientists, but still present a challenge. One crucial aspect in this respect is an adequate delineation of the sets of publications that determine the reference values to which a scientist's publication record and its citation impact should be compared. The structure of partition cells formed by intersecting fixed subject categories in a database has been proposed to approximate a scientist's specialty more closely than can be done with the broader subject categories. This paper investigates this cell structure's suitability as an underlying basis for methodologies to assess individual scientists, from two perspectives:
(1) Proximity to the actual structure of publication records of individual scientists: The distribution and concentration of publications over the highly fragmented structure of partition cells are examined for a sample of ERC grantees;
(2) Proximity to customary levels of accuracy: Differences in commonly used reference values (mean expected number of citations per publication, and threshold number of citations for highly cited publications) between adjacent partition cells are compared to differences in two other dimensions: successive publication years and successive citation window lengths.
Findings from both perspectives are in support of partition cells rather than the larger subject categories as a journal based structure on which to construct and apply methodologies for the assessment of highly specialized publication records such as those of individual scientists.

**Introduction**
Leading excellent researchers have become the focus of an increasing number of funding programs worldwide, in a context of growing global competition to recruit the best. Grantees are selected by peer review, generally accepted as the primary methodology to evaluate research quality, but nevertheless under pressure from criticisms. These include issues related to the methodology itself as well as to resources (over-sollicitation of experts) and to workload (high numbers of applications to be evaluated in a same round). Quantitative indicators that are a good proxy for quality as perceived by peers can be valuable complements to peer reviews in evaluation procedures. Advanced bibliometric indicators in particular have been applied at the levels of research groups, university departments and institutes (van Raan, 2005). Still, several decades after the introduction of the Web of Science in its earliest form, it remains a challenge to develop bibliometric indicators that can be adequately applied to publication records as 'small' and specialized as those of individual scientists. Recently multiple conference sessions have been explicitly dedicated to methodological and ethical aspects of individual-level evaluative bibliometrics (14th Conference of the International Society for Scientometrics and Informetrics, Vienna; 18th

---

International Conference on Science and Technology Indicators, Berlin; 5th Biennial Atlanta Conference on Science and Innovation Policy, Atlanta). Among efforts to develop novel and better adapted methodologies for the assessment of individual scientists was the partition based field normalization method, using cells smaller than subject categories as reference sets to more closely fit highly specialized publication records (Rons, 2012). These partition cells and the larger subject categories both are journal-based structures. Among methods to distinguish between fields (Schubert & Braun, 1996), also paper-based methodologies can produce reference sets representing scientific specialties. Depending on the methodology these may however be available for certain disciplines only, or may require intensive calculation or manual efforts. One way to proceed is the arrangement of individual papers in fine-grained classification schemes maintained by certain research domains, e.g. in the Chemical Abstracts database used in a bibliometric analysis by Neuhaus and Daniel (2009). Another is to use advanced algorithms to generate approximations of specialties, involving e.g. bibliographic coupling (Kessler, 1963), co-citation (Small, 1973), direct citation, co-words (Callon et al., 1983), or a combination. The journal-based and paper-based methodologies offer different advantages. In particular, paper-based structures can offer high precision, and journal-based structures high stability. Partition cells, as journal-based structures more subdivided than subject categories, combine higher precision with stability. Directly determined by the fixed subject categories, the partition cells are available as a more subdivided basis for all disciplines and for various methodologies. Besides the normalization context in which they were originally proposed, they can for instance also be applied in the identification of highly cited publications (Rons, 2013). Progress in domain delimitation however responds to just one of the challenges faced in comparative bibliometric assessments of individual scientists. Other issues include data accuracy, approaches to interdisciplinary research and to multidisciplinary and general journals in journal based reference standards, and the variation in bibliometric characteristics among equally distinguished scientists. A methodology needs to cope with these various issues to be able to produce results that are strongly correlated with peer judgments on individual scientists. Such results could mean a significant support in evaluation procedures, as they would help evaluators concentrate on candidates in a crucial range or dedicate more of their time to the most complex dossiers, and thus perform their tasks more efficiently. Gaging the actual improvements in accuracy contributed by using partition cells rather than subject categories in various methodologies requires further investigation. This paper focuses on the ability to more closely fit reference sets to publication records of individual scientists, and on the level of accuracy attained as compared to customary levels for other variables.

**Data and Methodology**
The investigations were conducted using publication and citation data from Thomson Reuters' online Web of Science (WoS), and for article type documents only (i.e. no multiple types) given their prominent role and status in the process of knowledge creation and dissemination. The investigated partition cells are formed by the fixed structure of overlapping WoS subject categories, such that each cell contains all publications associated to exactly the same combination of subject categories (Rons, 2012). These cells form reference sets of an intermediate size between journals (relatively narrow) and entire subject categories (relatively wide). Reference sets at different levels of aggregation generate different reference values to which a performance can be compared. The level of aggregation for instance strongly influences which publications belong to the top-cited groups (Zitt, Ramanana-Rahary & Bassecoulard, 2005). Broad discipline-based systems (such as the 60 subfields developed by the groups in Leuven and Budapest; Glänzel & Schubert, 2003) were found to be a sufficiently accurate basis for assessments at country and institution level (Glänzel et al., 2009). Compared to these levels, individual scientists have much more specialized research profiles. The structure of partition cells, introduced in the specific context of highly specialized publication records, offers a more subdivided basis of reference sets that are neither so wide as to include strongly diverging publication and citation practices from unrelated areas, nor so narrowly fit as to reflect a researcher's potential bias towards certain publication media.

*ERC Grantee Publication Records*
From the WoS, all articles were collected that were authored by a sample of grantees from the first European Research Council's ERC Advanced Grants call (2008). The ERC Advanced Grants are aimed at exceptional, established, scientifically independent research leaders with a track-record of significant research achievements in the last 10 years (http://erc.europa.eu/advanced-grants). The selection is based on international peer review with excellence as the sole criterion, using a system of 25 discipline-based panels of high-level scientists. For this bibliometric investigation, grantees were observed from two ERC panels representing domains with strongly different publication productivity and citation characteristics:
- 21 grantees in the panel 'Mathematical foundations' (slower characteristics);
- 14 grantees in the panel 'Fundamental constituents of matter' (faster characteristics).
For the identification of the grantees' articles among those of homonyms, cv-information was used next to the data available in the WoS. For each grantee, the distribution of articles over partition cells was calculated in the 8-year publication period preceding the call (2000-2007). The mean number of citations per article, in a 5-year citation window for each publication year, was calculated per grantee as a rough indication of citation levels (a more thorough investigation of citations in view of indicator design or verification being beyond the scope of this paper).

*Levels of Accuracy*
It is customary good practice to compare observed citation-based values for a studied entity to reference values for the same publication year and citation window. Differences between reference values in subsequent publication years or citation window lengths, indicate the influence on results of using a reference time frame just one year off the appropriate one. Similarly, differences can be observed between reference values in adjacent partition cells associated to a same subject category, indicating the influence on results of using a reference research area just next to the appropriate one. These differences and their relative magnitudes were studied for two commonly used reference values:
- The mean expected number of citations per article ($e$), used in the general standard field normalized citation rates (Braun & Glänzel, 1990; Moed et al., 1995);
- The threshold number of citations for outstandingly cited articles ($T$) as determined by the methodology of Characteristic Scores and Scales (Glänzel & Schubert, 1988), used for the identification of highly cited publications, which have since long been regarded as bibliometric emanations of research excellence at the level of individual scientists (Garfield, 1986).
Differences between pairs of reference values were calculated in three dimensions:
(1) In subsequent publication years (pairs 2005-2006 and 2006-2007);
(2) In subsequent citation window lengths (pairs 3-4 and 4-5 years, including the publication year);
(3) In adjacent partition cells $C$ (pairs $C_i$-$C_{i,j}$ and $C_{i,j}$-$C_j$, where $C_i$, $C_{i,j}$ and $C_j$ contain the articles in journals assigned respectively to subject categories $i$ only, $i$ and $j$ combined, and $j$ only.

The investigations were conducted in two domains with strongly different citation characteristics related to the two ERC panels in the 'ERC Grantee Publication Records' section above, for the following cells:
- $C_M$, $C_{M;MA}$ and $C_{MA}$ in the domain of Mathematics, where M and MA respectively stand for the subject categories 'Mathematics' and 'Mathematics Applied'.
- $C_{AA}$, $C_{AA;PPF}$ and $C_{PPF}$ in a sub-domain of physics, where AA and PPF respectively stand for the subject categories 'Astronomy & Astrophysics' and 'Physics, Particles & Fields'.
For each of the two reference values $e$ and $T$, 36 pair-wise comparisons of positive reference values $x$ and $y$ were made in each of the three dimensions, calculating the absolute relative difference $r_e$ and $r_T$ as $r=2*|x-y|/(x+y)$.

**Results and Discussion**
*ERC Grantee Publication Records*
Figures 1 and 2 show the distribution and concentration of articles in the observed 8-year publication period for each grantee in the two panels, indicating the shares of articles per cell. For each grantee the top share is highlighted. The cells listed are all those that contain articles by at least one grantee (excluding for instance 2% of the subject category 'Mathematics' in Figure 1, and 13% of the subject category 'Physics, Particles & Fields' in Figure 2). Within the highly fragmented structure of cells, the grantees' articles are strongly concentrated in one or a few cells, and totally absent or very limitedly present in the other cells that are associated to the same subject categories. The partition cells therefore offer a more accurate structural basis than the larger subject categories to delimit the publication environment in which to position these scientists' performances. For instance, some grantees publish the top share of their articles in a cell associated to exactly one subject category, and no articles in any of the other cells associated to that subject category ($M_1$, $M_2$, $M_3$, $M_{13}$, $M_{15}$). One grantee publishes the top share of his articles in a cell associated to a combination of two subject categories, and only 3% in all other cells associated to either of these two subject categories ($F_3$).

Figures 1 and 2 further demonstrate a strong diversity in publication profiles among grantees who were evaluated in a same disciplinary panel. Even grantees who have been co-authors, presumably at least partly working on closely related topics, may publish their top shares of articles in different cells ($F_1$ and $F_9$). For some grantees who do have highly similar distributions of articles over cells, productivities and citation levels differ by a factor 2 ($M_1$ and $M_2$; $M_6$ and $M_7$). Similar observations of publication records with very different characteristics for equally distinguished scientists in a same discipline were made by Sugimoto and Cronin (2012) for six information scientists. Among the factors that play a role in such variation may be contextual issues regarding the individual scientist (local environment, personal choices), and issues regarding the organization of scientific literature in the scientist's domain (e.g. the presence of interdisciplinary or multispecialty journals or database categories).

Figure 1: Distribution of articles over partition cells for grantees from the ERC Advanced Grant call 2008, panel 'Mathematical foundations'.

| Grantee: | $M_1$ | $M_2$ | $M_3$ | $M_4$ | $M_5$ | $M_6$ | $M_7$ | $M_8$ | $M_9$ | $M_{10}$ | $M_{11}$ | $M_{12}$ | $M_{13}$ | $M_{14}$ | $M_{15}$ | $M_{16}$ | $M_{17}$ | $M_{18}$ | $M_{19}$ | $M_{20}$ | $M_{21}$ |
|---|---|---|---|---|---|---|---|---|---|---|---|---|---|---|---|---|---|---|---|---|---|
| Number of articles 2000-2007: | 11 | 6 | 6 | 22 | 25 | 20 | 11 | 14 | 20 | 10 | 27 | 19 | 12 | 98 | 7 | 19 | 62 | 20 | 15 | 22 | 31 |
| Cell $C_X$ with X: | Share of articles per cell | | | | | | | | | | | | | | | | | | | | |
| ACS;CSAI;MA | | | | | | | | | | | | | | | | 5% | | | | | |
| ACS;MA | | | | | | | | | | | | 11% | | | | | | | 20% | | 3% |
| BRM;BAM;CSIA;MCB;SP | | | | | | | | | | | | | 1% | | | | | | | | |
| B | | | | | | | | | | | | | | | | | 2% | | | | |
| B;MCB | | | | | | | | | | | | | | | | | 24% | | | | |
| Bp;EB | | | | | | | | | 4% | | | | | | | | | | | | |
| BF;E;MIA;SSMM | | | | | | | | | | | | | | | | | | | 7% | | 6% |
| CCS;RS;S | | | | | | | | | 4% | | | | | | | | | | | | |
| CB | | | | | | | | | | | | | | | | | 2% | | | | |
| CSAI | | | | | | | | | | | | | | | | | | | | 23% | |
| CSAI;CSC;EEE | | | | | | | | | | | | | | | | | | | | 5% | |
| CSAI;CSIA;EB;RNMMI | | | | | | | | | | | | | | | | | | | | 9% | |
| CSAI;EEE | | | | | | | | | | | | | | | | | | | | 9% | |
| CSHA;CSIS;CSSE;CSTM | | | | | | | | 5% | | | | | | | | | | | | | |
| CSIS | | | | | | | | | | | | | 2% | | | | | | | | |
| CSIS;EEE | | | | | | | | | | | | | 3% | | | | | | | | |
| CSIA;CSSE;MA | | | | | | | | | | | | | | | | | 11% | | | | |
| CSIA;EB;EEE;ISPT;RNMMI | | | | | | | | | | | | | | | | | | | | 5% | |
| CSIA;MA | | | | | | | | | | | | | | | | | 2% | | | | |
| CSIA;MIA;Mc;PFP | | | | | | | | | | | | | | | | 5% | | | | | |
| CSIA;Mc | | | | | | | | | | | | | | | | 11% | | | | | |
| CSIA;PM | | | | | | | | | | | | | | | | 37% | | 10% | | 5% | |
| CSSE;MA | | | | | | | | | | | | | 1% | | | | | | | | |
| CSSE;MA;M | | | | | | 5% | | | | | | | | 8% | 6% | | | | | | |
| CSSE;ORMS;MA | | | | | | 5% | | | | | | | | | | | | | | | |
| CSTM | | | | | | | | | | | | | 2% | | | | | | | | |
| CSTM;M | | | | | | | | | | | | | 1% | | | | | | | | |
| CSTM;MA | | | | | | | | | | | | | 6% | | | | | | | | |
| CSTM;MA;L | | | | | | | 5% | | | | | | 1% | | | | | | | | |
| CSTM;MA;M | | | | | | | | | | | | | | | | | | 5% | | 5% | |
| CSTM;M;SP | | | | | | | | | 15% | | | | 13% | | | | | | | | |
| E;MIA;SSMM | | | | | | | | | | | | | | | | | | | | | 3% |
| EB;RNMMI | | | | | | | | | | | | | | | | | | | | 9% | |
| EEE;O;ISPT | | | | | | | | | | | | | | | | | | | | 5% | |
| EMd;MA | | | | | | | | | | | 4% | | | | | | | | | | |
| EMd;MIA | | | | | | | | | | | 4% | | | | | | | | | | |
| EMd;MIA;Mc | | | | | | | | | | | 19% | | | | | | | | | | |
| HM | | | | | | | | | | | 4% | | | | | | | | | | |
| MCB | | | | | | | | | | | | | | | | | 2% | | | | |
| M | 100% | 100% | 100% | 95% | 88% | 85% | 82% | 21% | 60% | | 4% | 53% | | 48% | 29% | 6% | | 20% | | | 16% |
| MA | | | | | | | | | | | | 56% | 11% | | 10% | | 26% | 32% | 30% | 13% | 14% | 6% |
| MA;M | | | | 5% | 4% | 5% | 9% | | 5% | 20% | | 16% | | 3% | 29% | | 8% | 5% | 27% | | 6% |
| MA;Mc | | | | | | | | | | | | | | | | | 2% | | | | |
| MA;PM | | | | | | | | | | | | | | | | | | 5% | | | 3% |
| MA;PM;SP | | | | | | | | | | | | | | | | | | | | | 13% |
| MA;PMd;PM | | | | | | | | | | | | | | | | | | 10% | | | |
| MA;SP | | | | | | | | | | | | | | | | | | | | | 16% |
| MIA | | | | | | | | | | | | | | | | | 2% | | | | |
| MIA;Mc | | | | | | | | | | | | 11% | | | | | | | | | |
| MIA;PM | | | | | | | | | 4% | | | | | | | 5% | | 10% | | | |
| MIA;PMd;PM | | | | | | | | | | | | | | | | | 2% | | | | |
| Mc | | | | | | | | | | | | | | | | 5% | | | | | |
| Mc;PFP | | | | | | | | | | | | | | | | | | 5% | | | |
| MS | | | | | 8% | | | | | | | | | | | | 2% | | | | 6% |
| N | | | | | | | | | | | | | | | | | | | | 5% | |
| N;Ni;RNMMI | | | | | | | | | | | | | | | | | | | | 9% | |
| Oc;CN | | | | | | | | | | | | | | | | | 2% | | | | |
| ORMS;MA | | | | | | | | | | | | | | | | | | | | | 3% |
| PFP;PM | | | | | | | | | | | | | | | | | 2% | | | 5% | |
| PM | | | | | | 10% | 9% | 71% | | 60% | | | | 33% | | 43% | 2% | | 5% | | |
| PMd | | | | | | | | | | 10% | | | | | | | | | | | |
| PMd;PM | | | | | | | | | | | | | | 8% | | | | | | | |
| PMd;PPF;PM | | | | | | | | 7% | | 10% | | | | | | | | | | | |
| RNMMI | | | | | | | | | | | | | 1% | | | | | | | | |
| SP | | | | | | | | | | | | | | 50% | 1% | | | | 10% | 13% | 16% |
| T;EM;Mc | | | | | | | | | | | | | | | | 5% | | | | | |
| | Mean number of citations per article in a 5-year citation window (including the publication year): | | | | | | | | | | | | | | | | | | | | |
| | 6 | 11 | 7 | 6 | 4 | 4 | 11 | 9 | 8 | 9 | 8 | 6 | 6 | 5 | 12 | 16 | 8 | 9 | 8 | 19 | 7 |

Subject categories:
ACS: Automation & Control Systems
B: Biology
BAM: Biotechnology & Applied Microbiology
BF: Business, Finance
Bp: Biophysics
BRM: Biochemical Research Methods
CB: Cell Biology
CCS: Cardiac & Cardiovascular Systems
CN: Clinical Neurology
CSAI: Computer Science, Artificial Intelligence
CSC: Computer Science, Cybernetics
CSHA: Computer Science, Hardware & Architecture
CSIA: Computer Science, Interdisciplinary Applications
CSIS: Computer Science, Information Systems
CSSE: Computer Science, Software Engineering

CSTM: Computer Science, Theory & Methods
E: Economics
EB: Engineering, Biomedical
EEE: Engineering, Electrical & Electronic
EM: Engineering, Mechanical
EMd: Engineering, Multidisciplinary
HM: Humanities, Multidisciplinary
ISPT: Imaging Science & Photographic Technology
L: Logic
M: Mathematics
MA: Mathematics, Applied
Mc: Mechanics
MCB: Mathematical & Computational Biology
MIA: Mathematics, Interdisciplinary Applications
MS: Multidisciplinary Sciences

N: Neurosciences
Ni: Neuroimaging
O: Optics
Oc: Oncology
ORMS: Operations Research & Management Science
PFP: Physics, Fluids & Plasmas
PM: Physics, Mathematical
PMd: Physics, Multidisciplinary
PPF: Physics, Particles & Fields
RNMMI: Radiology, Nuclear Medicine & Medical Imaging
RS: Respiratory System
S: Surgery
SP: Statistics & Probability
SSMM: Social Sciences, Mathematical Methods
T: Thermodynamics

*Data sourced from Thomson Reuters Web of Knowledge (formerly referred to as ISI Web of Science). Web of Science (WoS) accessed online 04&09.10.2013.*

Figure 2: Distribution of articles over partition cells for grantees from the ERC Advanced Grant call 2008, panel 'Fundamental constituents of matter'.

| Grantee: | $F_1$ | $F_2$ | $F_3$ | $F_4$ | $F_5$ | $F_6$ | $F_7$ | $F_8$ | $F_9$ | $F_{10}$ | $F_{11}$ | $F_{12}$ | $F_{13}$ | $F_{14}$ |
|---|---|---|---|---|---|---|---|---|---|---|---|---|---|---|
| Number of articles 2000-2007: | 34 | 19 | 117 | 64 | 85 | 98 | 38 | 29 | 23 | 33 | 149 | 48 | 69 | 50 |
| Cell $C_X$ with X: | Share of articles per cell | | | | | | | | | | | | | |
| AA | | | | | | | | 7% | | | | | | |
| AA;PMd | | 5% | | | | | | 7% | 4% | | | | | |
| AA;PMd;PPF | 3% | | | 3% | | | | | | | | | | |
| AA;PPF | | 63% | | 8% | | | | | 26% | | | | | |
| Bp | | | | | | | | | | | | 2% | | |
| CA | | | | | | | | | | | 1% | | | |
| CMd | | | | | | | | | | | | | 1% | |
| CMd;CP;NN;MSM;PA;PCM | | | | | | | | | | | 6% | | | |
| CP;MSM;PA;PCM | | | | | 1% | | | | | | | | | |
| CP;NN;MSM | | | | | | | | | | | | 2% | | |
| CP;NST;PAMC | | | | | | | | | | | | | 1% | |
| CP;PAMC | | | | | | | | | | 6% | | | | 4% |
| CSHA;CSIS;O;Tc | | | | 1% | | | | | | | | | | |
| CSSE;MA | | | | | | | | | | | 1% | | | |
| CSTM;EEE;MA | | | | | | | | | | | 1% | | | |
| CSTM;PPF;PM | | | 2% | | | | | | | | 5% | | | |
| ESD;PMd | | | | | 1% | | | | | | 1% | | | |
| EEE | | | | | | | | | | | 1% | | | |
| EEE;II | | | | | | | | 3% | | | | | | |
| EEE;O;PA | | | | | 1% | 1% | | | | | 5% | | | 2% |
| EEE;O;Tc | | | | | | | | | | | 3% | | | |
| EMd;II | | | | 1% | | | | | | | | | | 2% |
| II;PA | | | | | 1% | | | | | | 1% | | 4% | |
| MSM;PA | | | | | | | | | | | | 4% | 1% | |
| MA | | | | | | 1% | | | | | | | | |
| MA;PM | | | | 2% | | | | | | | | | | |
| MS | 3% | | 1% | | 18% | 6% | 21% | 3% | 4% | | 3% | 10% | 6% | 4% |
| O | | | 3% | | 12% | 46% | 11% | | | 18% | 13% | 31% | 4% | 24% |
| O;PA | | | | 2% | | | 8% | | 3% | 6% | 1% | 2% | 1% | 12% |
| O;PAMC | | | 50% | | 14% | 2% | 16% | 31% | | 36% | 26% | | | 18% |
| PA | | | | | 2% | 2% | | | | | 2% | 17% | 12% | 10% |
| PAMC | | | | | | | | | 9% | | | | | |
| PCM | | | 1% | | | | | | | | 8% | | | 6% |
| PFP | | | | | | | | | | | | 29% | | |
| PFP;PM | | | | | | 13% | | | | | | 7% | | |
| PFP;PN | | | | | | | | | | | | 3% | | |
| PM | | | | 8% | | | | | | | 1% | | | |
| PMd | 21% | 21% | 42% | 30% | 47% | 28% | 42% | 45% | 39% | 24% | 32% | 17% | 25% | 18% |
| PMd;PM | | | 1% | | | | 3% | | | | 4% | | | |
| PN;PPF | | | | 2% | | | | 4% | | | | | 1% | |
| PPF | 74% | 11% | | 48% | | | | | 22% | | | | | |
| RNMMI | | | | | | | | | | | | | 3% | |
| T;CP | | | | | | | | | | | 1% | | | |
| Mean number of citations per article in a 5-year citation window (including the publication year): | | | | | | | | | | | | | | |
| 23 | 47 | 25 | 22 | 48 | 34 | 52 | 89 | 61 | 22 | 25 | 71 | 32 | 31 |

Subject categories:
AA: Astronomy & Astrophysics
Bp: Biophysics
CA: Chemistry, Analytical
CMd: Chemistry, Multidisciplinary
CP: Chemistry, Physical
CSHA: Computer Science, Hardware & Architecture
CSIS: Computer Science, Information Systems
CSSE: Computer Science, Software Engineering
CSTM: Computer Science, Theory & Methods
EEE: Engineering, Electrical & Electronic
EMd: Engineering, Multidisciplinary
ESD: Education, Scientific Disciplines
II: Instruments & Instrumentation
MA: Mathematics, Applied
MS: Multidisciplinary Sciences
MSM: Materials Science, Multidisciplinary
NN: Nanoscience & Nanotechnology
NST: Nuclear Science & Technology
O: Optics
PA: Physics, Applied
PAMC: Physics, Atomic, Molecular & Chemical
PCM: Physics, Condensed Matter
PFP: Physics, Fluids & Plasmas
PM: Physics, Mathematical
PMd: Physics, Multidisciplinary
PN: Physics, Nuclear
PPF: Physics, Particles & Fields
RNMMI: Radiology, Nuclear Medicine & Medical Imaging
T: Thermodynamics
Tc: Telecommunications

*Data sourced from Thomson Reuters Web of Knowledge (formerly referred to as ISI Web of Science).*
*Web of Science (WoS) accessed online 21.10.2013.*

*Levels of Accuracy*

Table 1 shows the reference values $e$ and $T$ in the observed range in each dimension (publication years, citation window lengths and partition cells), and Figure 3 the absolute relative differences $r_e$ and $r_T$ between pairs of reference values in each dimension (successive publication years, successive citation window lengths and adjacent partition cells). The observed variation of reference values presents an example of the magnitudes that can be attained, and in particular of the relative magnitudes in the three dimensions. For the investigated suite of cells and time frame, the pair-wise differences between reference values are of a same order of magnitude in the three dimensions, with the differences for adjacent partition cells ($r_e$: 0-42%; $r_T$: 0-36%) lying higher than the differences for successive publication years ($r_e$: 0-18%; $r_T$: 0-31%) and lower than the differences for successive citation window lengths ($r_e$: 18-51%; $r_T$: 8-59%). A comparison to reference values per partition cell is therefore of comparable importance in terms of potential error generated, as a comparison to reference values per publication year and for the exact citation window observed, which is customary.

For publication records of large entities such as institutes or countries, calculating reference values per larger subject category has been experienced to generate sufficiently accurate global results, and is common practice. Such publication records are widely spread out over numerous partition cells. In such conditions, compensating effects from publications in cells 'advantaged' and 'disadvantaged' by reference values calculated at a more global level are likely to occur, and limit the overall influence on results of less accurate reference values for certain sub-domains. Figures 1 and 2 illustrate that publication records of individual scientists are concentrated in only a limited number of partition cells. In such conditions the same compensating effects are unlikely to occur, and it is equally recommendable to use reference values calculated per partition cell, as it is to use reference values calculated per publication year and for the exact citation window observed.

Table 1. Reference values in the observed range of publication years, citation window lengths and partition cells.

| Cell $C_X$ with X: | Publication year | Number of articles | Reference values | | | | | |
|---|---|---|---|---|---|---|---|---|
| | | | Mean expected number of citations per article ($e$) | | | Threshold number of citations for outstandingly cited articles ($T$) | | |
| | | | Citation window length (years, incl. publication year) | | | | | |
| | | | 3 | 4 | 5 | 3 | 4 | 5 |
| M | 2005 | 10055 | 1.2 | 2.0 | 2.8 | 6.1 | 8.3 | 11.8 |
| | 2006 | 10400 | 1.4 | 2.2 | 3.0 | 6.4 | 11.4 | 15.3 |
| | 2007 | 11410 | 1.4 | 2.2 | 2.9 | 6.5 | 11.6 | 12.5 |
| M;MA | 2005 | 4322 | 1.3 | 2.2 | 3.2 | 6.3 | 11.6 | 15.9 |
| | 2006 | 4531 | 1.6 | 2.6 | 3.6 | 8.1 | 12.5 | 16.7 |
| | 2007 | 5717 | 1.8 | 2.9 | 3.9 | 8.4 | 12.4 | 18.0 |
| MA | 2005 | 4589 | 2.0 | 3.3 | 4.7 | 8.2 | 14.7 | 20.1 |
| | 2006 | 4871 | 2.1 | 3.5 | 4.8 | 9.6 | 14.9 | 20.1 |
| | 2007 | 5288 | 2.3 | 3.7 | 5.0 | 11.8 | 17.7 | 24.6 |
| AA | 2005 | 8203 | 9.4 | 13.4 | 17.3 | 37.0 | 52.9 | 69.8 |
| | 2006 | 8798 | 9.8 | 14.2 | 18.0 | 37.8 | 58.0 | 74.0 |
| | 2007 | 8920 | 9.8 | 13.9 | 17.6 | 38.3 | 54.6 | 70.0 |
| AA;PPF | 2005 | 2398 | 10.3 | 14.2 | 17.7 | 39.7 | 56.3 | 69.8 |
| | 2006 | 2574 | 10.1 | 14.2 | 17.6 | 36.5 | 52.0 | 64.7 |
| | 2007 | 2545 | 10.9 | 14.8 | 18.1 | 39.5 | 55.5 | 68.9 |
| PPF | 2005 | 1780 | 9.0 | 12.1 | 14.6 | 38.7 | 54.6 | 67.5 |
| | 2006 | 1728 | 8.4 | 11.1 | 13.4 | 32.8 | 45.7 | 55.1 |
| | 2007 | 1815 | 9.1 | 12.0 | 14.6 | 34.1 | 45.3 | 57.5 |

*Data sourced from Thomson Reuters Web of Knowledge (formerly referred to as ISI Web of Science). Web of Science (WoS) accessed online 10.01.2013-30.05.2013.*

Figure 3: Absolute relative differences $r_e$ and $r_T$ between pairs of reference values $e$ and $T$ in three dimensions.

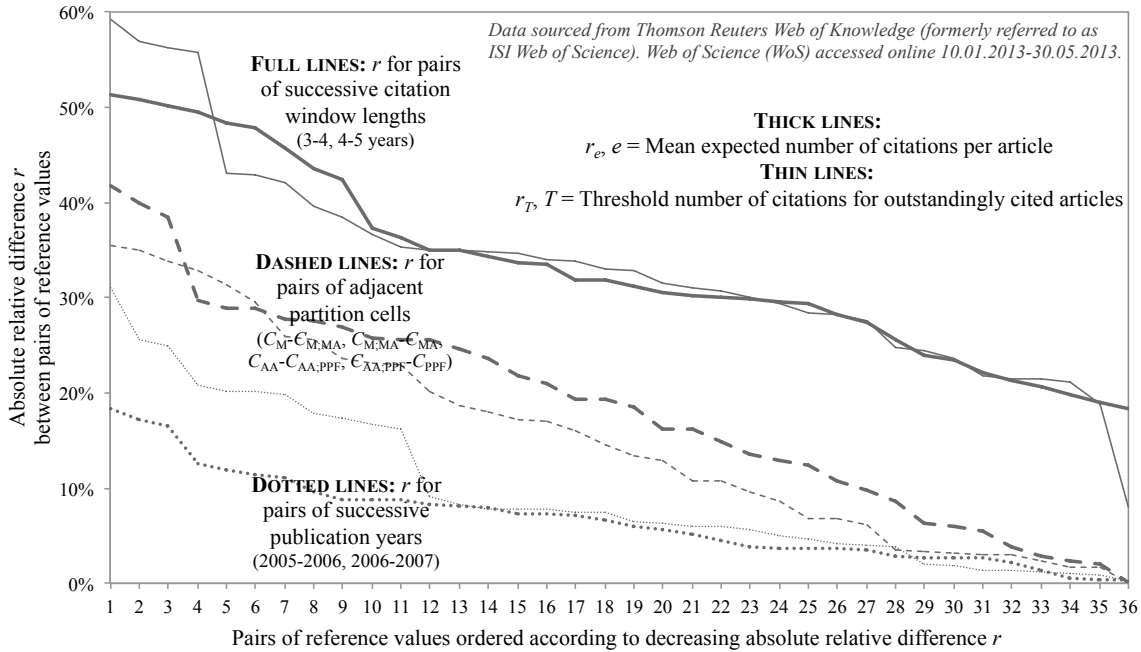

**Conclusion**

The results of the investigation show that (at least for the examined domains, sample of scientists and range of bibliometric variables):
- Reference sets of publications can be more closely fit to a scientist's specialty when using partition cells than when using the larger subject categories;
- The level of accuracy attained when using partition cells is comparable to customary levels of accuracy applied when calculating reference values per publication year and per citation window length.

These findings are in support of partition cells rather than the larger subject categories as a basis for methodologies for the assessment of highly specialized publication records. As more subdivided structures directly determined by the fixed subject categories, partition cells combine higher precision with stability, and are available for all disciplines. The actual effect in particular cases of applying partition cells rather than subject categories can be expected to vary with specialty and indicator, such study however being beyond the scope of the present paper. Reported examples of important differences between expected citation rates calculated for WoS subject categories versus smaller paper-by-paper classification structures indicate that the effect can be decisive (differing by a factor 2.2 for a sub-domain of neurology reported on by Bornmann et al., 2008, and by a factor 1.7 for a sub-domain of biochemistry reported on by Neuhaus & Daniel, 2009). Similar comparisons between results of different indicators based on subject categories versus partition cells thus are an important element for further studies investigating actual improvements associated to the latter. The outcomes are not only of interest in a context of assessments of individual scientists, but may also be relevant for other applications involving domain delineation and reference sets for publication records in scientific specialties, regardless of the volume of publications and the number of authors concerned.